\documentclass[times,review,3p]{elsarticle}

\usepackage{framed,multirow}

\usepackage{amssymb}
\usepackage{amsmath}

\usepackage{latexsym}
\usepackage{subfigure}
\usepackage{multirow}
\usepackage{booktabs}
\usepackage{hyperref}
\usepackage{lineno}

\usepackage{url}
\usepackage{xcolor}

\usepackage{tikz}
\usepackage{catchfilebetweentags}
\usetikzlibrary{arrows.meta, shapes.geometric, calc, shadows}

\definecolor{newcolor}{rgb}{.8,.349,.1}

\journal{Journal of Computational Physics}

\begin{document}
\begin{frontmatter}

\title{Some improvements on Moment-of-Fluid method in 3D rectangular hexahedrons}%
\author[1,2]{Zhouteng Ye}
\ead{yzt9zju@gmail.com}
\author[2]{Mark Sussman}
\ead{sussman@math.fsu.edu}
\author[1]{Xizeng Zhao\corref{cor1}}

\cortext[cor1]{Corresponding author: 
  email: xizengzhao@zju.edu.cn;  
  }

\address[1]{
Ocean College, Zhejiang University, Zhoushan 316021, Zhejiang, People’s Republic of China}
\address[2]{
Department of Applied \& Computational Mathematics, Florida State University, United States}

\cortext[cor1]{Corresponding author: 
  email: xizengzhao@zju.edu.cn;  
  }


\begin{abstract}

The moment-of-fluid method (MOF) is an extension of the volume-of-fluid method with piecewise linear interface construction (VOF-PLIC). 
In  MOF reconstruction, 
the optimized normal vector is determined from the reference centroid and the volume fraction by iteration.
The state-of-art work by \citet{milcent_moment--fluid_2020} proposed an analytic gradient of the objective function, 
which greatly reduces the computational cost.
In this study, 
we further accelerate the MOF reconstruction algorithm by using Gauss-Newton iteration instead of Broyden-Fletcher-Goldfarb-Shanno (BFGS) iteration.
We also propose an improved initial guess for MOF reconstruction, 
which improves the efficiency and the robustness of the MOF reconstruction algorithm.
Our implementation of the code and test cases are available on our Github repository.
\end{abstract}


\end{frontmatter}



\section{Introduction}

A lot of scientific and engineering problems involve with tracking the interface between different materials.
Multiple volume tracking/capturing methods, 
such as volume-of-fluid method (VOF) \cite{hirt_volume_1981}, 
level set method \cite{osher_fronts_1988}, 
front tracking method \cite{unverdi_front-tracking_1992}
have been introduced to describe the motion of interface explicitly or implicitly.
Among those methods, the volume-of-fluid method with piecewise line interface construction (VOF-PLIC) is one 
of the most widely used methods in tracking the interface within the Eulerian framework.

Conventional VOF-PLIC reconstructs the normal vector of the reconstructed interface
by using the a stencil that contains the information of the neighbouring grids,
for example, Parker and Youngs' algorithm \cite{parker_two_1992},
mixed Youngs-centered algorithm (MYC) \cite{aulisa_interface_2007},
and the efficient least squares volume-of-fluid interface reconstruction algorithm (ELVIRA) \cite{pilliod_second-order_2004}.
Although some of the VOF-PLIC reconstruction algorithms are second-order accuracy,
when there is not enough information from the neighbouring grids, 
for example, very small scale droplets,
VOF-PLIC algorithm may not reconstruct the interface accurately.

Moment of Fluid method (MOF) \cite{dyadechko_moment--fluid_2005,dyadechko_reconstruction_2008}
introduces the centroid as the additional constraint to determine the normal vector of the reconstruction plane.
When there is  no data from adjacent cells used in reconstruction,
MOF reconstruction resolves the the interface with a smaller minimum scale than the VOF-PLIC algorithm.
In MOF reconstruction,
evaluating the objective function and its partial derivative is the most expensive part during iteration.
The original MOF algorithm by \citet{dyadechko_moment--fluid_2005} has to call
a very complex polyhedra intersection algorithm for 5 times in every iteration.
Although later study by \citet{chen_improved_2016} reduces the calling of the geometric algorithm
to one time each iteration, the computational cost is still heavy.
\citet{lemoine_moment--fluid_2017} made their first attempt to derive an analytic form of that describes 
the objective function as the minimum distance from the reference centroid to a closed, continuous curve
in 2D Cartesian grid.
This is a fully 2D MOF algorithm as solution to the objective function can be obtained by 
computing the cubic or quartic roots of polynomials instead of iteration.
Unfortunately, this approach cannot be extended to 3D \cite{milcent_moment--fluid_2020}.
\citet{milcent_moment--fluid_2020} derived an analytic form of the partial derivative of objective function in 3D rectangular hexahedron,
by using the analytic form,
the computational cost is significantly reduced.
The algorithm could be more than 200 times faster than the conventional MOF reconstruction \cite{dyadechko_moment--fluid_2005}.

In this study, we further accelerates the MOF algorithm based on the analytic gradient by \citet{milcent_moment--fluid_2020}.
We use the Gauss-Newton algorithm instead of the Broyden-Fletcher-Goldfarb-Shanno (BFGS) algorithm in \citet{milcent_moment--fluid_2020}.
Although Gauss-Newton algorithm has been used in other MOF studies \cite{jemison_coupled_2013,asuri_mukundan_3d_2020},
no detailed comparison between the two algorithms has been carried out.
We show that the Gauss-Newton algorithm significantly reduces the number of gradient calculation.
We also propose an improved form of initial guess,
which provides a closer initial guess to the global minimum.
The improved initial guess helps to reduce the iteration step and improves the robustness
of the Gauss-Newton algorithm.

The run-time ratio and robustness of the method could be implementation-dependent.
Out implementation of the code and test cases are available on our Github repository (https://github.com/zhoutengye/NNMOF).
All the numerical tests are done on a workstation with Intel(R) Xeon(R) Platinum 8270 processors
with Intel compiler 2020.

\section{Moment of Fluid reconstruction}

\subsection{Description}
As an extented the VOF-PLIC method, 
the MOF method reconstructs the interface in a 3D rectangular hextahedron cell
with a plane
\begin{equation}
\label{eq:mofplane}
\left\{\mathbf{x} \in \mathbb{R}^{3} \mid \mathbf{n} \cdot\left(\mathbf{x}-\mathbf{x_0}\right)+\alpha=0\right\},
\end{equation}
where $\mathbf{n}$ is the normal vector, $\mathbf{x}$ is the reference point of the cell,
$\mathbf{x_0}$ is the origin of the Cartesian coordinate,
either the center of the cell or the lower corner of the cell, 
depending on the computational algorithm.
$\alpha$ is the parameter that represents the distances from the reference point $\mathbf{x}$.
The volume of the cutting polyhedron by the reconstruction plane satisfies
\begin{equation}
\label{eq:mofvolumeconstraint}
\left|F_{\mathrm{ref}}(\mathbf{n}, \alpha)-F_{A}(\mathbf{n}, \alpha)\right|=0.
\end{equation}
and
\begin{equation}
\label{eq:minimizecentroid}
E_{\mathrm{MOF}}=\left\|\mathbf{x}_{\mathrm{ref}}-\mathbf{x}_{A}(\mathbf{n}, b)\right\|_{2}
\end{equation}
In addition to the constraint on volume fraction,
the MOF reconstruction also minimizes error of the centroid with
\begin{equation}
\label{eq:mofminimizeangle}
E_{\mathrm{MOF}}\left(\Phi^{*}, \Theta^{*}\right)=\left\|\mathbf{f}\left(\Phi^{*}, \Theta^{*}\right)\right\|_{2}=
\min _{(\Phi, \Theta):\rm{Eq.} \eqref{eq:mofvolumeconstraint} \text { holds }}\|\mathbf{f}(\Phi, \Theta)\|_{2}
\end{equation}
Eq. \eqref{eq:mofminimizeangle} minimizes the distance in 3D with two parameters by converting the
normal vector in Eq. \eqref{eq:mofplane} to the polar angle and azimuthal angle in a spherical coordinate system.
Eq. \eqref{eq:mofminimizeangle} is a non-linear least square problem for $\Phi$ and $\Theta$,
it is solved with optimization algorithm via iteration.

\subsection{Optimization of the centroid}
We use the Gauss-Newton algorithm to minimize Eq. \eqref{eq:mofminimizeangle}. 
Starting with an initial guess of $\Phi_{0}, \Theta_{0}$, the solution procedure are:

1. Find the centroid $\mathbf{x}_k$ corresponds with the angle $(\Phi_{k}, \Theta_{k})$.

2. Determine the Jacobian matrix $\mathbf{J_k}$ using the analytic solution \citet{milcent_moment--fluid_2020}.

3. Determine the sift angle
\begin{equation}
\label{eq:shiftangle}
(\Delta \Phi_k, \Delta \Theta_k) = - \mathbf{H_k J_{k}^{T} f_k}
\end{equation}

where $\mathbf{H_k} = \mathbf{J_k^{T}{J_k}}$ is the Hessian matrix. 
In this problem, the dimension of the Hessian matrix is $2 \times 2$.

4. Update angle $(\Phi_{k+1},\Theta_{k+1}) = (\Phi_{k} + \Delta \Phi_k, \Theta_{k} + \Delta \Theta_k) $.

The iteration stops while convergence conditions are full-filled. 
Multiple convergence criteria can be adopted, for example, 
centroid error, 
error of the gradient of the objective function,
minimum advance angle, 
and maximum iteration step.

In this problem, 
even though the gradient of the objective function has been significantly boosted by the analytic gradient of \citet{milcent_moment--fluid_2020} compared with 
the conventional numerical gradient approach,
calculating the objective function $\mathbf{f}$ and the gradient objective function $(\frac{\partial \mathbf{f}}{\partial \Phi},\frac{\partial \mathbf{f}}{\partial \Theta}$) still takes most of the computational time during iteration.
The number of calling the gradient algorithm determines the total computational cost of the iteration.
In the original MOF method \cite{dyadechko_moment--fluid_2005}, the non-linear optimization Eq. \eqref{eq:mofminimizeangle} is solved with
Broyden-Fletcher-Goldfarb-Shanno (BFGS) algorithm, which is also used in \citet{chen_improved_2016,milcent_moment--fluid_2020}.
In BFGS algorithm, the advance angle $(\Delta \Phi_{k}, \Delta \Theta_{k})$ needs to be determined by a line search algorithm.
In the numerical tests of \citet{milcent_moment--fluid_2020}, 
every iteration needs 8-10 steps of line search,
which means the total number of calling the gradient algorithm is much bigger than the number of the iteration step.

The main advantage of the BFGS algorithm over Gauss-Newton algorithm is that the BFGS algorithm approximates the inverse of the
Hessian matrix,
which avoids the calculation of the inverse of the Hessian matrix directly.
However, in this problem, 
the shape of the Hessian matrix is only $2 \times 2$,
making the cost of the inverse matrix algorithm negligible.
While in Gauss-Newton algorithm, 
the shift angle is directly determined by Eq. \eqref{eq:shiftangle},
so that the number of calling the gradient algorithm equals to the iteration step.
Compared with BFGS algorithm, the number of calling the gradient algorithm is much smaller than the BFGS algorithm if
both algorithms converge with same iteration steps.

Other non-linear optimization could potentially be used in minimizing Eq. \eqref{eq:mofminimizeangle}. 
For example, the Levenberg-Manquardt algorithm \cite{madsen_methods_2004} is known as an inproved Gauss-Newton algorithm using a trust region approach.
Although the Levenberg Manquardt algorithm is more robust than the Gauss-Newton algorithm,
finding the value of the trust region involves trial computation of the objective function and its gradient, 
which could significantly increase the computational cost.
For efficiency, 
we use the Gauss-Newton algorithm other than Levenberg-Manquardt algorithm.
To ensure the robustness of the Gauss-Newton algorithm,
we provide an improved initial guess in next subsection.

\subsection{Initial guess of the normal vector}
The choice of initial guess is important because
there may exists multiple local minima in the objective function.
\citet{dyadechko_moment--fluid_2005} suggested the form
\begin{equation}
\label{eq:initialguess_1}
\mathbf{n_0^1} = \mathbf{x_c}(\Omega) - \mathbf{x_{ref}}
\end{equation}
as safe initial guess.
\citet{dyadechko_moment--fluid_2005} also claimed that 
the line-search algorithm guarantees the initial guess finally reaching the global minima.
In Gauss-Newton iteration, 
the step is automatically determined,
there is no trial-step selection.
The Gauss-Newton algorithm is more likely to be sensitive to the initial guess than 
BFGS with line search algorithm used in the study of \citet{dyadechko_moment--fluid_2005, milcent_moment--fluid_2020}.

\ExecuteMetaData[figures.tex]{fig:flood}

We propose a new form of the initial guess in this section. 
To better demonstrate the philosophy of our proposed initial guess, 
we simplify the 3D problem to 2D by setting the polar angles $\Phi=\pi/2$,
which simplifies the 3D problem to a 2D problem.
Fig. \ref{fig:flood} shows the locus of the centroids of the cutting polygon by a line interface in a unit cell.
The solid lines in Fig. \ref{fig:flood} (b) corresponds with the evolution of the centroid with a fixed azimuthal angle $\Phi$ (See Fig. \ref{fig:flood}(a)),
and the dashed lines in Fig. \ref{fig:flood}(b) corresponds with the evolution of the centroid with a fixed volume fraction $V$ (See Fig. \ref{fig:flood}(c)).
When the volume fraction $V>1/2$ (with red color in Fig. \ref{fig:flood}),
the centroid can be determined by finding the centroid of its symmetric cutting polygon.
We only discuss the case when $V>1/2$ in this section.
For any of the reference centroid  close to one of a vertex of the cell,
the corresponding centroid $\mathbf{x_0}$ of an initial guess $\Phi_0$ could be very close to the reference centroid $\mathbf{x_c}$,
but has a big error with the exact azimuthal angle $\Phi$.

We show the error of initial guess 
Eq. \eqref{eq:initialguess_1}
in Fig. \ref{fig:initialguess} (a). 
Eq. \eqref{eq:initialguess_1} gives a good initial guess in most areas except for the 
region that is near the face of the cell.
Those regions correspond with very small volume fraction. 
We propose another candidate initial guess by assuming the reference centroid 
as the centroid of a right triangle reconstruction (or a trirectangular tetrahedron in 3D).
\begin{equation}
\label{eq:initialguess_2}
\mathbf{n_0^2} = \mathbf{\frac{1}{\tilde x_v(\Omega) - \mathbf{x_{ref}}}},
\end{equation}
where  $\mathbf{\tilde{x}_{v}(\Omega)}$ is the vertex of the quadrant (or octant in 3D) that $\mathbf{{x}_{ref}}$ is located.
The error map of the azimuthal angle is plotted in Fig. \ref{fig:initialguess} (b).
The right triangle approximates the small volume correctly especially when the centroid is 
near one of the vortex of the grid cell.
We evaluates the value of the objective function with the two candidate initial guesses and
pick the one with smaller error of the centroted
\begin{equation}
\label{eq:initialguess_3}
\mathbf{n}_{0}=\underset{\left\{\mathbf{n}_{0}^{1}, \mathbf{n}_{0}^{2}\right\}}{\arg \min }\left\{E_{\mathrm{MOF}}\left(\mathbf{n}_{0}^{1}\right), E_{\mathrm{MOF}}\left(\mathbf{n}_{0}^{2}\right)\right\}
\end{equation}
The error of the azimuthal angle error $\Delta \Phi_{e}$ of Eq. \eqref{eq:initialguess_3} are plotted in Fig. \ref{fig:initialguess} (c).
In 2D case, the maximum error of the polar angle by Eq. \eqref{eq:initialguess_3} is approximately $\pi/20$,
while the maximum error of the polar angle from Eq. \eqref{eq:initialguess_1} is about $\frac{\pi}{4}$.
We also teste the initial guess in 3D.
The error of the initial guess $\Delta \Theta+\Delta \Phi$ by Eq. \eqref{eq:initialguess_1} is about $\frac{\pi}{2}$.
By using our improved initial guess, 
the error reduces to about $\frac{\pi}{5}$.

\ExecuteMetaData[figures.tex]{fig:initialguess}

\section{Numerical tests}
\subsection{Reconstruction test}

\ExecuteMetaData[figures.tex]{fig:distribution}

In this section, we test the accuracy and robustness of our MOF reconstruction with Gauss-Newton algorithm with improved initial guess.
Two criteria are evaluated: the CPU time and the robustness. 
Three data sets are generated by finding the exact centroid of 
a cutting polyhedron of a unit cube by a plane.
We use data-sets with different distribution to show the performance of our algorithm, especially the robustness for extreme cases (See Fig. \ref{fig:distribution}):
(1) Exponential case with a normal distribution;
(2) Uniform case corresponds with uniform distribution;
(3) Extreme case with a shifted normal distribution which contains more values near 0 or 1.
In this test, 
the tolerance for iteration is $10^{-8}$, 
the maximum iteration step is 100.

\ExecuteMetaData[tables.tex]{tab:error}

\ExecuteMetaData[tables.tex]{tab:time}

In Table \ref{tab:Error},
the error of the Gauss-Newton algorithm with original initial guess increases 
when more extreme data appears in the test data set,
while the BFGS shows a better robustness than the Gauss-Newton algorithm.
With the improved initial guess, 
both algorithms show a very good robustness in all test cases.
In BFGS algorithm,
each iteration needs a line search algorithm to determine the shift angle,
which needs to call the gradient algorithm for multiple times.
While in Gauss-Newton iteration,
the shift angle is automatically determined in each iteration,
the gradient algorithm only has to be called for once.
In Table \ref{tab:time}, 
it is observed that the averaged iteration step using Gauss Newton algorithm is smaller than 
that in BFGS algorithm.
When taking the line search into account,
the number of calling gradient algorithm in BFGS algorithm is about 5 times larger than that 
in Gauss Newton algorithm.
The Gauss Newton algorithm is about 3 times faster than the BFGS algorithm with analytic reconstruction.
It should note that we also compared out algorithm with the conventional MOF reconstruction \citep{dyadechko_moment--fluid_2005}, 
our  algorithm is more than 1000 faster than the conventional MOF reconstruction.

\subsection{Advection test}
In the previous test,
the optimized centroid is consistent with the reference centroid.
However, 
in practical, the optimized centroid may not be consistent with the reference  centroid.
In order to test the accuracy and robustness of the proposed algorithm with non-linear reconstruction,
we test our algorithm with a 3D Zalesak's problem \citet{enright_hybrid_2002}.
We use a directional splitting Lagrangian Explicit scheme for the advection of volume fraction \citep{aulisa_interface_2007} and updates the centroid by calculating the evolution of corresponding Lagrangian centroid.
For advection of the volume fraction and centroid,
We use a directional splitting Lagrangian Explicit scheme similar with the VOF-PLIC advection in \citet{aulisa_interface_2007}. 

The difference between different method are very small (With an $L_1$ error of $O^{-7}$),
which shows the robustness of our algorithm. 
The averaged time of iteration and computational cost are listed in Table \ref{tab:Zalesak}.
With our improved initial guess, 
the averaged number of iteration decreases in Gauss-Newton algorithm and BFGS algorithm.
The Gauss-Newton algorithm with the improved initial guess gets an acceleration of 3
to the algorithm of the \cite{milcent_moment--fluid_2020}.

\ExecuteMetaData[tables.tex]{tab:zalesak}

\section{Conclusions}
In this study, we show that using Gauss-Newton algorithm instead of BFGS algorithm significantly helps to accelerate the iteration in MOF reconstruction.
We also proposed an improved initial guess which makes the Gauss-Newton iteration more robust.
Our improved initial guess along with the Gauss-Newton algorithm is about 4 times faster 
than the BFGS algorithm by \citet{milcent_moment--fluid_2020} in reconstruction
and about 2 times faster in advection test.

\section{Acknowledgments}
The support provided by  National Science Foundation of China (Grant Nos. 51979245, 51679212) and 
China Scholarship Council (CSC) and the during a visit of Zhouteng Ye to Florida State University is acknowledged.

\bibliographystyle{model1-num-names}
\bibliography{refs}

\end{document}